\documentclass[english,aps,preprint]{revtex4}
\usepackage[T1]{fontenc}
\usepackage[latin9]{inputenc}

\usepackage{babel}

\begin{document}

\title{Comment on ``Event Excess in the MiniBooNE Search for $\bar{\nu}_{\mu}\rightarrow\bar{\nu}_{e}$
Oscillations''}

\author{M. SAHIN}

\email{m.sahin@etu.edu.tr}

\affiliation{TOBB University of Economics and Technology, Physics Division, Ankara,
Turkey}

\author{S. SULTANSOY}

\email{ssultansoy@etu.edu.tr}

\affiliation{TOBB University of Economics and Technology, Physics Division, Ankara,
Turkey}

\affiliation{Institute of Physics, National Academy of Sciences, Baku, Azerbaijan}

\author{S. TURKOZ}

\email{turkoz@science.ankara.edu.tr}

\affiliation{Ankara University, Department of Physics, Ankara, Turkey}
\maketitle

In a recent Letter, Aguilar-Arevalo \emph{et al}. \cite{Aguilar}
has claimed that {}``An excess of $20.9\pm14.0$ events is observed
in the energy range $475<E_{\nu}^{QE}<1250$ MeV, which, when constrained
by the observed $\bar{\nu}_{\mu}$ events, has a probability for consistency
with the background-only hypothesis of $0.5\%$''. To the contrary
we show the wrongness of this statement. Indeed, the data given in
the Table II, shows that the deviation between observed data and background
in the energy range $475<E_{\nu}^{QE}<1250$ MeV is approximately
$1\sigma$, which corresponds to probability $\sim32\%$ \cite{PDG}
for background-only hypothesis. Actually, the above mention statement
of the paper is correct for the energy range $475<E_{\nu}^{QE}<675$
MeV (see corresponding row of the Table II).

At the first glance one may assume that this mistake reflects
an ordinary misprint only. Unfortunately the detailed analysis of
the paper shows that this not the case (see paragraph started with
{}``Figure 1 (top) shows...'' on the page 3 as well as the last
paragraph of the paper). Fortunately the above mentioned mistake does
not change results given in the rest of the paper, which contains
interesting results.


\begin{thebibliography}{2}
\bibitem{Aguilar} A. Aguilar-Arevalo \emph{et al., }Phys. Rev. Lett.\emph{
}\textbf{105,}\textbf{\emph{ }}181801 (2010).

\bibitem{PDG} K. Nakamura \emph{et al}., (Particle Data Group), J.
Phys. G \textbf{37}, 075021 (2010).
\end{thebibliography}
\end{document}